\documentclass{article}
\usepackage{arxiv}
\usepackage{url}            
\usepackage{amsfonts}       
\usepackage{nicefrac}       
\usepackage{microtype}      
\usepackage{hyperref}
\usepackage{cite}
\usepackage{xcolor}
\usepackage{comment}
\usepackage{multirow}
\usepackage{multicol}
\usepackage{booktabs}
\usepackage[T1]{fontenc}
\usepackage{lipsum}
\usepackage{algorithm}
\usepackage{algpseudocode}
\usepackage[utf8]{inputenc}
\usepackage{array}
\usepackage{lettrine}
\usepackage{graphicx}
\usepackage{stfloats}
\usepackage{amsmath}
\usepackage{amssymb}
\graphicspath{ {./images/} }

\title{Fractional-order SEIQRDP model for simulating the dynamics of COVID-19 epidemic}

\author{
 Mohamed A. Bahloul \\
  Computer, Electrical and Mathematical Sciences and Engineering Division (CEMSE)\\ King Abdullah University of Science and Technology (KAUST)\\ Thuwal 23955-6900, Makkah Province, Saudi Arabia\\
  \texttt{mohamad.bahloul@kaust.edu.sa} \\
   \And
 Abderrazak Chahid \\
    Computer, Electrical and Mathematical Sciences and Engineering Division (CEMSE)\\ King Abdullah University of Science and Technology (KAUST)\\ Thuwal 23955-6900, Makkah Province, Saudi Arabia\\
  \texttt{abderrazak.chahid@kaust.edu.sa} \\
  \And
 Taous-Meriem Laleg-Kirati \\
  Computer, Electrical and Mathematical Sciences and Engineering Division (CEMSE)\\ King Abdullah University of Science and Technology (KAUST)\\ Thuwal 23955-6900, Makkah Province, Saudi Arabia\\
  \texttt{taousmeriem.laleg@kaust.edu.sa} \\
}
\begin{document}
\maketitle
\begin{abstract}
The novel corona-virus disease (COVID-19), known as the causative virus of outbreak pneumonia initially recognized in the mainland of China, late December 2019. COVID-19 reaches out to many countries in the world, and the number of daily cases continues to increase rapidly. In order to simulate, track, and forecast the trend of the virus spread, several mathematical and statistical models have been developed. \textit{Susceptible-Exposed-Infected-Quarantined-Recovered-Death-Insusceptible (SEIQRDP)} model is one of the most promising dynamic systems that has been proposed for estimating the transmissibility of the COVID-19. In the present study, we propose a Fractional-order SEIQRDP model to analyze the  COVID-19 epidemic. The Fractional-order paradigm offers a flexible, appropriate, and reliable framework for pandemic growth characterization. In fact, fractional-order operator is not local and consider the memory of the variables. Hence, it takes into account the sub-diffusion process of confirmed and recovered cases growth. The results of the validation of the model using real COVID-19 data are presented, and the pertinence of the proposed model to analyze, understand and predict the epidemic is discussed.
\\
\end{abstract}
\section{Introduction}
COVID-19 is a respiratory disease caused by the new coronavirus that was first identified in Wuhan, Hubei province, China, late December 2019 \cite{huang2020clinical}. This novel virus soon began to spread out around the world, and on 30 January, WHO declared the outbreak a Public Health Emergency of International Concern (PHEIC). On 11 March, WHO Director-General marked COVID-19 as a pandemic \cite{bedford2020covid,guo2020origin}. The transmission of COVID-19 is primarily through respiratory droplets and contact routes \cite{liu2020community}. Accordingly, ideal interventions to control the spread include: quarantine, isolation, increase home confinement, promoting the wearing of face masks, travel restrictions, the closing of public space, and cancellation of events. The number of cases increased rapidly to more than 3.25 million cases, including around 231,000 deaths worldwide as of April, $30$, 2020.

The novel corona-virus caused-pneumonia has attracted the attention of scientists with different backgrounds ranged from epidemiology to data science, mathematics, and statistics. Accordingly, several studies based on either statistics or mathematical modeling have been proposed for better analysis and a deep understanding of the evolution of this epidemic. Besides, enormous efforts have been devoted to predicting the inflection point and ending time of this epidemic in order to help make decisions concerning the different measures that have been taken by different governments.

Mathematical models have always played a crucial role in the understanding of the spread of the virus and in providing relevant guidelines for controlling the pandemic. Basically, mathematical paradigms are considered to be very useful in this context, as they provide detailed mechanisms for the epidemic dynamics. Among the most widely investigated model for the characterization of the COVID-19 outbreak in the world is the classic \textit{Susceptible-Exposed-Infectious-Recovered} (SEIR) model \cite{huang2020clinical}. Since the outbreak of the virus, the SEIR model has been intensively utilized to evaluate the effectiveness of multiple measures, which seems to be a challenging task for general other estimation methods \cite{tang2020updated,tang2020estimation,shen2020lockdown}. For instance, it has been employed to evaluate the effects of lock-down on the transmission dynamics between provinces in China, such as the effect of the lock-down in Hubei province on the transmission in Wuhan and Beijing \cite{shen2020lockdown}. In addition, a cascading scheme of the SEIR model has been studied to emulate the process of transmission from infection sources to humans. This approach was efficient to reach useful conclusions on the outbreak dynamics\cite{chen2020mathematical}. The work of \cite{peng2020epidemic} presented a generalization of the classical SEIR known as \textit{Susceptible-Exposed-Infected-Quarantined-Recovered-Death-Insusceptible} (SEIQRDP) model, for the epidemic analysis of COVID-19 in China. This generalization is based on the introduction of a new quarantined state and of the effect of preventive actions, which are considered as crucial epidemic parameters for COVID-19, such as the latent and quarantine time. As a result, considering this generalization of SEIR, the estimation of the inflection point, ending time, and total infected cases in widely affected regions was accurately determined and verified. 

Over the last decade, the fractional-order derivative (FD), defined as a generalization of the conventional integer derivative to a non-integer order (arbitrary order) operator, has been used to simulate many phenomena involving memory and delays including epidemic behavior \cite{gonzalez2014fractional,shaikh2020mathematical}. FD models offer a promising tool for the description of complex systems, in addition to their potential to incorporate accurately the memory and delay involved in the systems, it also provides more flexibility than classical integer-order models in fitting the data accurately \cite{magin2006fractional}.In this paper, we investigate a fractional version of the  (\textit{SEIQRDP}) model in modeling the  COVID-19 epidemic's dynamics. We will use real data to fit the model and analyze the results, and we will provide some insights on the interpretation and role of the fractional derivatives  \cite{shaikh2020mathematical}.

This paper is organized as follows. In Section II, we will recall some basic concepts from the $SEIQRDP$ epidemic model and the fractional-order derivatives. Section III is devoted to the presentation of fractional-order (\textit{SEIQRDP}) epidemic model (\textit{F-SEIQRDP}). In Section IV, we present the materials and methods. Section v present the estimation results. The last section discusses the obtained results and provides some future directions on the use of the model for analyzing and controlling the COVID-19 epidemic.  
\section{Preliminaries}
\begin{figure*}[b!]
    \centering
    \includegraphics[height=10cm,width=12cm]{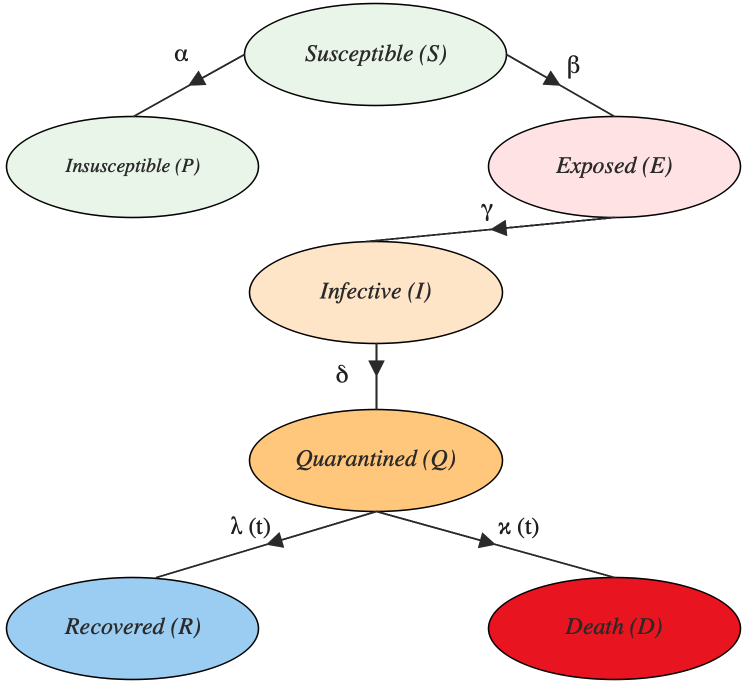}
    \caption{Flowchart illustrates the COVID-19 based SEIQRDP epidemic model, adopted from Fig.1 in \cite{peng2020epidemic}.}
    \label{fig:fig1}
\end{figure*}
In this section, we recall some basic concepts from the \textit{SEIQRDP} epidemic model and fractional-order derivatives theory.

\subsection{Generalized Epidemic Model: SEIQRDP}
As described in \cite{peng2020epidemic} the \textit{SEIQRDP} model consists of seven sub-populations (states), \textit{i.e} $\{S(t),E(t),I(t),Q(t),R(t),D(t),P(t)\}$ denoting at time $t$ the followings:
\begin{itemize}
    \item $S(t)$: Susceptible cases. 
    \item $E(t)$: Exposed cases, which are infected but not yet be infectious, in a latent period.
    \item $I(t)$: Infectious cases, which have infectious capacity but not yet be quarantined.
    \item $Q(t)$: Quarantined cases,  which are confirmed and infected.
    \item $R(t)$: Recovered cases.
    \item $D(t)$: Dead cases.
    \item $P(t)$: Insusceptible cases.
\end{itemize}
It contains also six parameters defined as follows:
\begin{itemize}
    \item $\alpha$: the protection rate.
    \item $\beta$: the infection rate.
    \item $\gamma$: the inverse of the average latent time.
    \item $\delta$: the rate at which infectious people enter in quarantine.
    \item $\lambda(t)$: a time-dependant coefficient used in the description of the cure rate. It is expressed as:
    \begin{equation}
        \lambda(t) =\lambda_0 \left[1-e^{-\lambda_1t}\right],
    \end{equation}
     where $\lambda_0$ and $\lambda_1$ are empirical coefficients.
    \item $\kappa(t)$: time-dependant coefficient used in the description of the mortality rate. It is expressed as:
    \begin{equation}
    \kappa (t) =\kappa_0 \ e^{-\kappa_1 t},
    \end{equation}
 where $\kappa_0$ and $\kappa_1$ are empirical coefficients.
\end{itemize}
It is worth to note that the time-dependent expressions of the cure rate, $\lambda(t)$, and the mortality rate, $\kappa(t)$, are assumed to be in the above forms based on the analysis of real data collected in some provinces in China, in January 2020, \cite{cheynet2020generalized}. The plot and analysis of this data showed the gradually increase of the cure rate and the quick decrease of the mortality rate. Furthermore, this assumptions are very reasonable by nature as the function of death rate in such pandemic always converges to zero while the cure rate continues increasing toward a consistent level. The other parameters are assumed to be constant as they are not fluctuating over time. 

The above parameters are controlled by the application of the preventive interventions as well as the effectiveness of the health systems in the investigated region. Figure 1 illustrates the relations between all the states. The dynamic of each state  is mathematically characterized by ordinary differential equations (ODE) as follows:\\
\begin{equation}
\left\{\begin{matrix}
\dfrac{\mathrm{d}S(t)}{\mathrm{d}t} = -\alpha S(t) -\beta \dfrac{S(t)I(t)}{N} \ \ \ \ \ \ \ \ \ \ \ \ \  \\ \\
\dfrac{\mathrm{d}E(t)}{\mathrm{d}t} = -\gamma E(t) +\beta \dfrac{S(t)I(t)}{N} \ \ \ \ \ \ \ \ \ \ \ \ \  \\ \\
\dfrac{\mathrm{d}I(t)}{\mathrm{d}t} = \gamma E(t) -\delta I(t) \ \ \ \ \ \ \ \ \ \ \ \ \  \ \ \ \ \ \ \ \ \ \\ \\
\dfrac{\mathrm{d}Q(t)}{\mathrm{d}t} = \delta I(t) -\lambda(t)Q(t) - \kappa(t) Q(t)  \ \  \\ \\
\dfrac{\mathrm{d}R(t)}{\mathrm{d}t} = \lambda(t) Q(t) \ \ \ \ \ \ \ \ \ \ \ \ \ \ \ \ \ \ \ \ \ \ \ \ \  \\ \\
\dfrac{\mathrm{d}D(t)}{\mathrm{d}t} = \kappa(t) Q(t) \ \ \ \ \ \ \ \ \ \ \ \ \ \ \ \ \ \ \ \ \ \ \ \  \\ \\
\dfrac{\mathrm{d}P(t)}{\mathrm{d}t} = \alpha S(t) \ \ \ \ \ \ \ \ \ \ \ \ \  \ \ \ \ \ \ \ \ \ \ \ \ \  
\end{matrix}\right.
\end{equation}
where $N$ represents the total population in the studied region expressed as $N = S+E+I+Q+R+D+P$. 
Comparing to the classical \textit{SEIR} model, \textit{SEIQRDP} is augmented by three new states, \{\textit{Q(t)}, \textit{D(t)} \& \textit{P(t)}\}. This new quarantined state \textit{Q(t)} and the recovery state \textit{R(t)} constitute, originally, the recovery state of the classical \textit{SEIR} model.
\subsection{Fractional-order Derivative}
In the past few decades, the theory of fractional calculus (FC) has gained significant research attention in several fields such as biology and epidemic modeling \cite{ahmed2007fractional,area2015fractional,islam2020integer}. This is originated from the interdisciplinary nature of this field as well as the flexibility and effectiveness of FC in describing complex physical systems. For example, the characterization of bio-impedance, modeling of the viscoelasticity and biological cells, and representing the mechanical properties of the arterial system, as well as respiratory systems, have been investigated extensively through the exploring of FC \cite{magin2006fractional,bahloul2018arterial,hilfer2000applications,bahloul2020assessment}. The concept of FC is not new dating from the pioneer conversation between \textit{L'Hopital} and \textit{Leibniz} in $1695$ that yielded to the generalization of the conventional integer derivative to a non-integer order operator \cite{podlubny1998fractional}, as follows:
 \begin{equation}
 D^{q}_{t}=
 \left\{ \begin{matrix}\dfrac{\mathrm{d^{q}} }{\mathrm{d} t^{\alpha}} 
 & \mbox{if} & q>0 \\ \\
 1,& \mbox{if} & q=0,\\ \\
 \int_{t}^{0}\left ( df \right )^{-q}& \mbox{if} & q<0
 \end{matrix}\right.
 \end{equation}
where ${q \in \mathbb{R}}$ is the order of the operator known as the fractional-order, and ${df}$ is the derivative function.

There are several fractional-order derivative definitions. In this work, we introduce the three most frequently used ones in the sense of the \textit{Riemann–Liouville}, \textit{Caputo} and \textit{Grünwald–Letnikov} FD-based definitions \cite{petras2011fractional,podlubny1999introduction, miller1993introduction}. The \textit{Grünwald–Letnikov} scheme based on finite differences has been adopted in the numerical implementation of the proposed \textit{F-GESIR}.

For a function $g(t)$ that satisfies some smoothness conditions then:\\
\begin{itemize}
    \item The \textit{Riemann–Liouville} definition is given as:
\end{itemize}
     \begin{equation}
 _a^{RL}D_t^{q}g\left ( t \right )=\dfrac{1}{\Gamma \left ( n-q  \right )}\dfrac{\mathrm{d^n} }{\mathrm{d} t^n}\int_{0}^{t}\left ( 1-\tau  \right )^{-q-1+n } g(\tau)d\tau.
 \end{equation}
 \begin{itemize}
     \item The \textit{Caputo} definition for FD is expressed as follows:
 \end{itemize}
     \begin{equation}
 _a^CD_t^{q}g\left ( t \right )=\dfrac{1}{\Gamma \left ( n-q  \right )}\int_{0}^{t}\left ( 1-\tau  \right )^{-q-1+n } \dfrac{\mathrm{d^n} }{\mathrm{d} t^n}g\left ( \tau  \right )d\tau, 
 \end{equation}
 where $\Gamma$ is the Euler gamma function and ($n-1<q<n$).
  \begin{itemize}
      \item The GL definition is given as:
  \end{itemize}
     \begin{equation}
 _a^{GL}D_t^{q}g\left ( t \right )=\lim_{h\to 0 }h^{-q}\sum_{j=0}^{\left[\frac{t-a}{h}\right] }(-1)^j\binom{q}{j}g(t-jh),
 \end{equation}
 where a is the terminal point and [.] means the integer part.\\
 
\section{Fractional-order SEIQRDP Epidemiological Model (F-SEIQRDP)}
 Similar to the \textit{SEIQRDP}, \cite{cheynet2020generalized}, the \textit{F-SEIQRDP} epidemiological model considers that the total population ($N$) is divided into seven sub-populations \textit{i.e} $\{S(t),E(t),I(t),Q(t),R(t),D(t),P(t)\}$. As the fractional-order derivative takes into account the history of the state, we believe that this operator is more suitable to describe the dynamics of the epidemic COVID-19. Using the definition of the fractional-order derivative operator, we consider that each state follows a fractional-order behavior.
 
 Considering the nonlinear FODEs in this matrix form:
 \begin{equation}
     D_t^q X(t)=A X(t)+L(X), 
 \end{equation}
 where, 
   \begin{equation*}
   D^q=[D^{q_S},D^{q_E},D^{q_I},D^{q_Q},D^{q_R},D^{q_D},D^{q_P}]^T 
 \end{equation*}
 is the fractional-order derivative operator for all the states and
 \begin{equation*}
   X=[S,E,I,Q,R,D,P]^T 
 \end{equation*}
 represents the state vector.
 \begin{equation*}
   A=  \begin{bmatrix}
-\alpha^{q_S} &      0 & 0 & 0 & 0 &  0& 0\\ 
 0& -\gamma^{q_E} & 0 &0  & 0 &  0&0 \\ 
 0& \gamma^{q_I} &  -\delta^{q_I}& 0 &  0&  0& 0\\ 
 0&  0&  \delta^{q_Q}& -\kappa^{q_Q}(t)-\lambda^{q_Q}(t) &  0& 0 & 0\\ 
 0& 0 &  0&\lambda^{q_R}(t)  & 0 & 0 &0 \\ 
 0& 0 & 0 & \kappa^{q_D}(t) & 0 & 0 &0 \\ 
 \alpha^{q_P}&  0& 0 & 0 &  0&0  &0 
\end{bmatrix}
 \end{equation*}
 represent the parameters. 
 
 $L(X)$ depicts the nonlinear term that is function of the susceptible and 
 \begin{equation*}
         L=S(t)  I(t) \begin{bmatrix}
-\dfrac{\beta^{q_S}}{N}\\
\\
-\dfrac{\beta^{q_E}}{N}\\ 
\\
0\\ 
0\\ 
0\\ 
0\\ 
0
\end{bmatrix}
 \end{equation*}
\section{Materials and Methods}
\subsection{Dataset}
The updated epidemic data of different countries around the world is collected from authoritative and known  sources as follows:
\begin{itemize}
\item \textbf{France}: The data is gathered from three main sources: "Agence Regionale de Sante", "Santé Publique France" and   "Geodes". This data is publicly available. \footnote{https://github.com/cedricguadalupe/FRANCE-COVID-19} 

\item \textbf{Italy}:  This data is provided by the  Italian government and it is publicly available.\footnote{ https://github.com/pcm-dpc/COVID-19} 

\item \textbf{Other countries}: The data is gathered from different official  sources: World Health Organization (WHO), Center of Disease Control and Prevention (CDC), the COVID Tracking Project (testing and hospitalizations), etc. The data repository is operated by the Johns Hopkins University Center for Systems Science and Engineering (JHU CSSE) and  Supported by ESRI Living Atlas Team and the Johns Hopkins University Applied Physics Lab (JHU APL). The repository is publicly available. \footnote{https://github.com/CSSEGISandData/COVID-19}
\end{itemize}

\subsection{Data fitting algorithm and numerical simulations}
The parameters of the proposed fractional-order model were estimated by a non-linear least square minimization routine, making use of the well-known $\mathrm{MATLAB-R2019b}$, function \textit{lsqnonlin}. This function is based on the trust-region reflective method \cite{coleman1996interior}. The steps used to obtain the optimal estimates are outlined in Algorithm 1.
\begin{algorithm}[!h]
\caption{Parameter estimation of epidemic data}
\begin{algorithmic}
\caption{Parameter estimation of epidemic data}
         \State \textbf{Input :} $t$:  Time in days
            \State $~~~~~~~~~R$: Recovered cases 
            \State $~~~~~~~~~I$: Confirmed cases 
            \State $~~~~~~~~~D$:  Dead cases 
            \State $~~~~~~~~~guess$:  The initial guess of the parameters  
            \State $~~~~~~~~~funmodel$:  The model to be fitted
\vspace{0.1cm}
         \State \textbf{Output:} $param$:  Fitted parameter of the  funmodel
\vspace{0.3cm}

\State  \textit{- Set the initial conditions}
       \State E = I; \Comment{Unknown but unlikely to be zero.}
       \State Q = I-R-D;
       \State input= [E; I; Q; R; D] 
       
\vspace{0.3cm}
\State  \textit{- Run the fitting optimization}
\State param=lsqcurvefit (t, \textit{funmodel}, \textit{guess}, input)
\end{algorithmic}
\end{algorithm}
\\
The fitting performances are evaluated using the followings metrics: 
\begin{equation}
 RMSE = \sqrt{\frac{1}{l}\Sigma_{i=1}^{l}{\Big({y(i) -\hat y(i)}\Big)^2}},
\end{equation}
and 
\begin{equation}
 ReMSE =  \frac{ \frac{1}{l} \Sigma_{i=1}^{l}{| {y(i) -\hat y(i)}|}}{max(y)},
\end{equation}
\noindent where $y$ and $\hat y$ are the real and fitted data, respectively. $l$ is the length of the data.
\begin{table*}[!t]
\centering
  \caption{ Estimation error comparison between the GESIR model and the proposed F-GESIR model for different countries. }
  \vspace{.5cm}
\label{tbl:table2} 
\resizebox{1\linewidth}{!}{
\renewcommand{\arraystretch}{2}
\begin{tabular}{|c|c|c|c|c|c|c|c|c|c|c|c|c|c|}
\hline
\multirow{2}{*}{\textbf{Model}}     & \multirow{2}{*}{\textbf{\begin{tabular}[c]{@{}c@{}}Estimation\\ Error\end{tabular}}} & \multicolumn{4}{c|}{\textbf{China}}&\multicolumn{4}{c|}{\textbf{Italy}} & \multicolumn{1}{c|}{\textbf{France}}  \\ \cline{3-11}   &   & \text{Beijing} & \text{Guangdong} & \text{Henan}  & \text{Hubei}    & \text{Lombardia} & \text{Veneto} & \text{\begin{tabular}[c]{@{}c@{}}Emilia-\\ Romagna\end{tabular}} & \text{Piemonte} & \text{\begin{tabular}[c]{@{}c@{}}Nouvelle-\\ Aquitaine\end{tabular}} \\ \hline
\multirow{2}{*}{\textbf{F-SEIQRDP}}   & $\mathrm{ReMSE}$   & 0.0625  & 0.0420    & 0.0215    & 0.0191& 0.0145    & 0.0071  & 0.0063 & {0.0048}& 0.0159   \\ \cline{2-11} 
 & $\mathrm{RMSE}$ & 22.17 & 53.26& 30.56  & 1677.89 & 648.37 & 97.21   & 108.912   & 97.96 & 63.94    \\ \hline
\multirow{2}{*}{\textbf{SEIQRDP}} & $\mathrm{ReMSE}$   & {0.1388}   & {0.1053}     & {0.0315}  & {0.0287}    & {0.0169}     & {0.0093}  & {0.0065} & {0.0047}    & {0.0231} \\ \cline{2-11} 
& $\mathrm{RMSE}$   & {45.84}  & {132.02}    &{45.51} & {2063.95} & {732.878}   &{119.08} & {114.67}   & {97.00}   & {85.44}                               
\\ \hline
\end{tabular}
}
\end{table*}
\section{Results}
\begin{figure*}[!b]
     \centering
     \includegraphics[height=14cm,width=16cm]{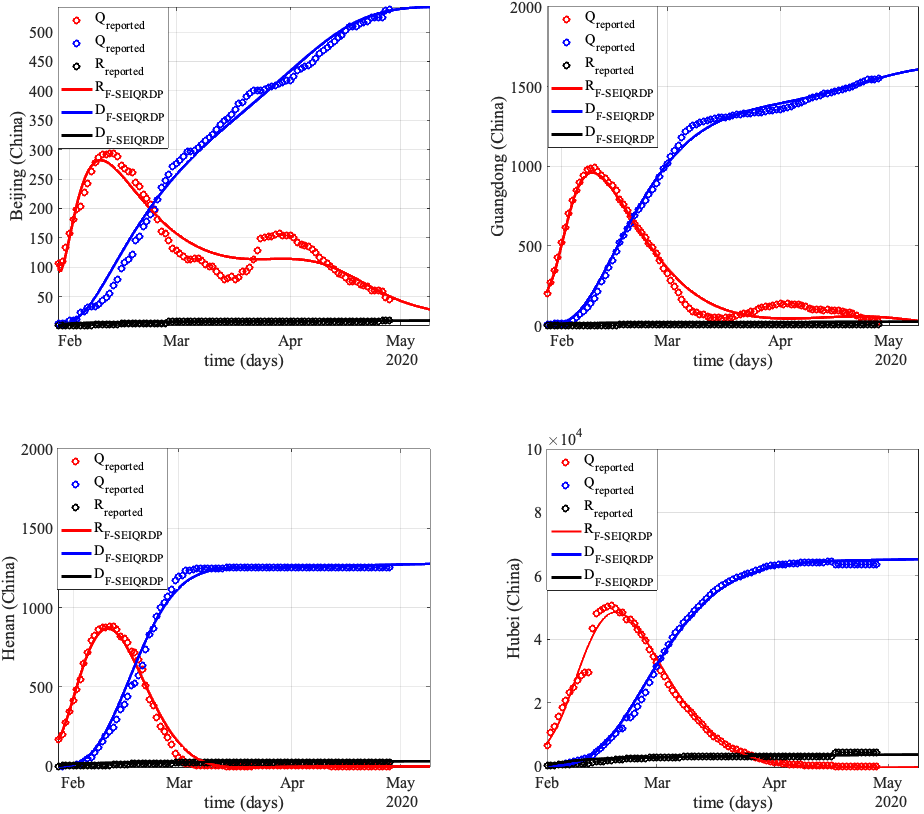}
     \caption{Predictions of the proposed fractional model using data from China.}\label{Fig:Data2}
\end{figure*}
The fitting performance of predicting the dynamics of $Q(T)$, $R(t)$, and $D(t)$ populations using \textit{SEIQRDP} and the proposed \textit{F-SEIQRDP} are presented in Table I. It is worth to note that \textit{SEIQRDP} epidemiological model can be considered as a special case of the \textit{F-SEIQRDP}, where all the fractional differentiation orders are equal to 1. From the reported results, it is clear that for all the studied cities the $\mathrm{ReRMSE}$ as well as the $\mathrm{RMSE}$ based on \textit{F-SEIQRDP} model are less than the ones reported using \textit{SEIQRDP}. These results show the usefulness of the fractional-order derivative operator in fitting real data of the pandemic. In addition, it demonstrates the potential of the fractional-order framework in estimating the size and the key milestones of the spread of the epidemic-COVID-19. The appropriateness of the fractional-order paradigm can be visualized from the fact that: FD operator is not local and depends on the strength of the memory that is controlled by the fractional differentiation order. On the other hand, the epidemiological dynamical process is involving the memory effect within the sub-diffusion process of confirmed and recovered cases growth.

\begin{figure*}[!t]
\vspace{-.5cm}
     \centering
     \includegraphics[height=14cm,width=16cm]{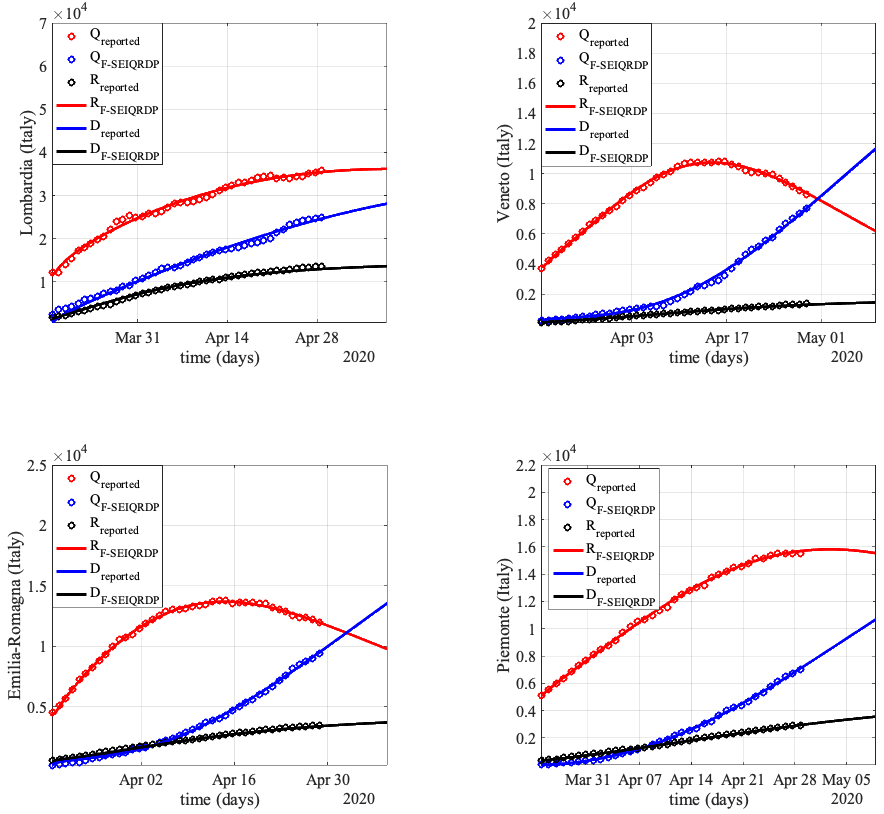}
     \caption{Predictions of the proposed fractional  model using data from Italy.}
     \label{Fig:Data1}
\end{figure*}
\begin{figure}[!b]
     \centering
     \includegraphics[height=6.5cm,width=8cm]{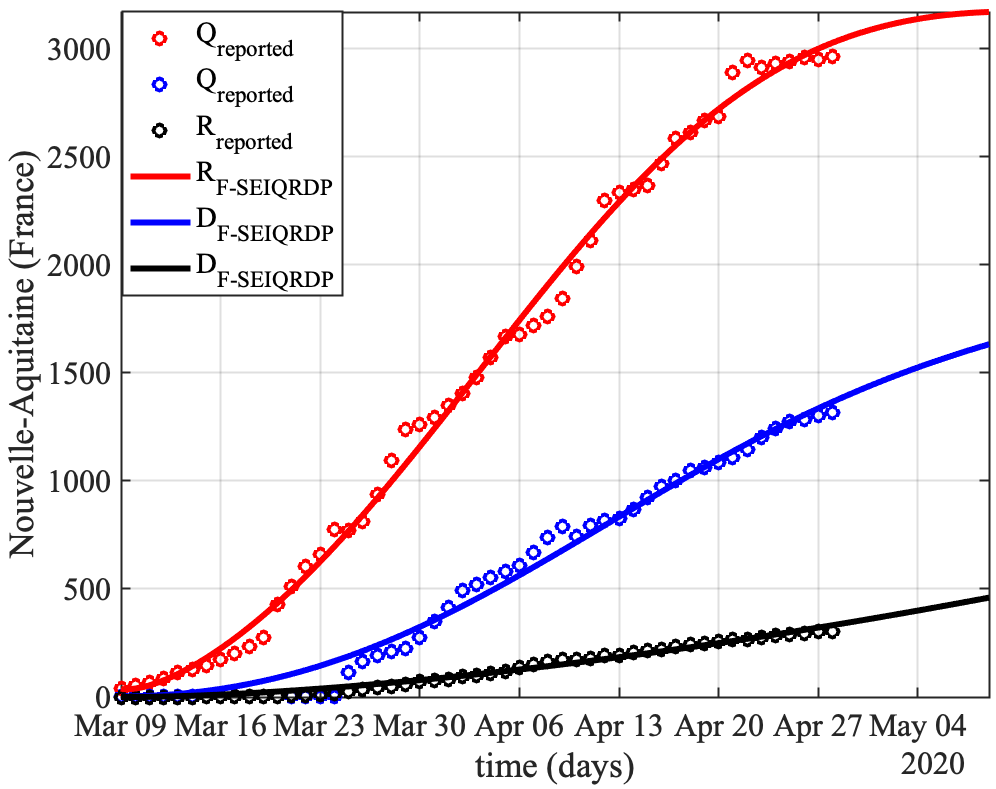}
     \caption{Predictions of the proposed fractional  model using  data from France.}
     \label{Fig:Data1}
\end{figure}
\begin{table*}[!b]
\centering
  \caption{Epidemic spread parameters estimation using the proposed F-SEIQRDP model for different countries.}
  \vspace{.5cm}
\label{tbl:table1} 
\resizebox{1\linewidth}{!}{
\renewcommand{\arraystretch}{2}
\begin{tabular}{|c|c|c|c|c|c|c|c|c|c|c|c|c|}
\hline
\textbf{Country}& \multicolumn{4}{c|}{\textbf{China}}& \multicolumn{4}{c|}{\textbf{Italy}}& \multicolumn{1}{c|}{\textbf{France}} \\ \hline
\text{\textbf{City}} & \textit{Beijing} & \textit{Guangdong} & \textit{Henan} & \textit{Hubei} & \textit{Lombardia} & \textit{Veneto} & \textit{\begin{tabular}[c]{@{}c@{}}Emilia-\\ Romagna\end{tabular}} & \textit{Piemonte} & \textit{\begin{tabular}[c]{@{}c@{}}Nouvelle\\ Aquitaine\end{tabular}} \\ \hline\hline
\text{{\begin{tabular}[c]{@{}c@{}}Population rate\\ (million)\end{tabular}}}          &21.54 &113.46 &94 & 58.50 & 10.04 & 4.90 & 4.45 & 4.37 & 5.98\\ \hline
\text{{\begin{tabular}[c]{@{}c@{}}Protection rate\\ $(\alpha)$\end{tabular}}}           &0.03820&0.03423&0.57635&0.23436&6.7E-5&0.05720&0.02336&0.02087&0.01251\\ \hline
\text{{\begin{tabular}[c]{@{}c@{}}Infection rate\\ $(\beta)$\end{tabular}}}             &0.00261&0.00105&0.00626&0.01940&0.8820&0.83078&0.72296&0.86569&0.50255\\ \hline
\text{{\begin{tabular}[c]{@{}c@{}}Latent time\\ $(\gamma^{-1})$\end{tabular}}}   &3.44862&2.92057&4.38832&4.88390&7.1748&7.34966&4.27373&8.34334&2.73742\\ \hline
\text{{\begin{tabular}[c]{@{}c@{}}Quarantine time\\ $(\delta^{-1})$\end{tabular}}}      &4.06448&4.76581&6.20915&6.59052&1.9342&1.84847&2.19720&2.15438&2.59576\\ \hline\hline
\textbf{$\lambda_0$}        &0.06655&0.12695&0.70887&0.33212&0.0393&0.32064&0.03188&0.09577&0.00390\\ \hline
\textbf{$\lambda_1$}        &0.05342&0.02932&0.00585&0.00623&0.0036&0.00361&0.07352&0.00582&0.00839\\ \hline
\textbf{$\kappa_0$}         &0.02674&0.05612&0.05950&0.05010&0.0342&0.00640&0.01320&0.01051&0.22037\\ \hline
\textbf{$\kappa_1$}         &0.68415&0.85788&0.67755&0.31469&0.0634&0.02783&0.03674&0.02526&0.37194\\ \hline\hline
\textbf{$q_S$}              &2.06&2.31&0.40&0.89&0.89&1.09&1.00&0.99&0.98 \\ \hline
\textbf{$q_E$}              &1.26&1.37&1.20&1.36&0.84&0.73&0.79&0.76&1.40 \\ \hline
\textbf{$q_I$}              &0.88&0.93&1.28&1.17&0.84&0.72&0.87&0.70&0.83 \\ \hline
\textbf{$q_Q$}              &0.87&1.04&1.03&1.06&0.86&0.94&0.88&0.94&0.92 \\ \hline
\textbf{$q_R$}              &0.87&0.99&1.04&1.05&0.23&1.14&1.30&0.71&0.24 \\ \hline
\textbf{$q_D$}              &1.20&1.20&1.20&1.20&1.20&1.20&1.0&1.0&2.59 \\ \hline
\textbf{$q_P$}              &1.00&1.00&1.00&1.00&1.00&1.00&.00&1.00&1.00 \\ \hline
\end{tabular}
}
\end{table*}
The parameter estimates for all the studied populations in different cities are reported in Table II. In this study, we choose different cities that present different circumstances in terms of the total number of population, the number of infected cases, and the lock-down schemes. Figures 2,3 and 4 show examples of the predicted dynamic of the quarantined, recovered and death sub-populations in \{Beijing, Guangdong, Henan, and Hubei\} cities in China, \{Lombardia, Veneto, Emilia-Romagna, and Piemonte\} cities in Italy, and   \{Nouvelle Aquitaine\} region in France, respectively. It is apparent in all the figures that we present the trend of the epidemiological dynamic till May $8^{th}$, which is the future concerning the date of simulation April, $29^{th}$. This shows the potential of the model in predicting the trend of the pandemic dynamic in the future. 

\section{Discussion and Conclusions}
Mathematical models are considered vital in every stage of the epidemic evolution. In fact, the simulation of the epidemic dynamics helps to track and monitor the spread of the virus. In addition, models are very effective in estimating the size of the pandemics and hence they might assist the specialized health actors to make the right decisions and minimize the losses. This paper proposes a new general fractional-order model for the evolution of the COVID-19 pandemic. The first validation results show the accurate model fitting using real COVID-19 data from different countries. The fractional-order derivatives provide new and pertinent parameters for the control of the epidemic. However, the model has some limitations that we can list as follows. 
\begin{itemize}
\item We observe that for countries with less data (reported cases in less than 30 days),  the estimation is less accurate because the trend does not appear yet (more than 28 days).

\item From country to country, the initial guess of parameters should be chosen carefully to guarantee the best possible fitting of the used optimization solver. 

\item The time-variable parameters need to be considered carefully because the countries do not have similar medical facilities and expertise, and they perform different numbers of tests per day. Besides, some countries have adopted some precautions strategies earlier than others such as quarantine, lock-down...etc.
\end{itemize}

\section*{Acknowledgment}
 \noindent Research reported in this publication was supported by King Abdullah University of Science and Technology (KAUST).
 
\section*{Funding}

\noindent This research project has been funded by King Abdullah University of Science and Technology (KAUST) Base Research Fund (BAS/1/1627-01-01).

\bibliographystyle{unsrt}  
\bibliography{references.bib}

\end{document}